\documentclass{article}
\usepackage{icrctc07,here}

\title{Testing the surface detector simulation for the Pierre Auger Observatory}
\shorttitle{Surface detector simulation
 for the Pierre Auger Observatory}

\authors{P.~L.~Ghia$^{1}$, for the Pierre Auger Collaboration$^{2}$ }
\shortauthors{Pierre Auger Collaboration}
\afiliations{$^1$ Istituto di Fisica dello Spazio Interplanetario, INAF, Torino and Laboratori Nazionali del Gran Sasso, INFN, Assergi, Italy\\$^2$ Pierre Auger Observatory, Av. San Mart{\'\i}n Norte 304, (5613) Malarg\"ue, Argentina}
\email{piera.ghia@lngs.infn.it}

\abstract{The building block of the surface detector of the Pierre Auger Observatory
 is a water Cherenkov tank. The response to shower particles is simulated using
 a dedicated program based on GEANT4. To check the simulation chain, we compare
 the simulated signals produced by cosmic muons at various zenith angles with experimental
 data from a special Cherenkov detector equipped with a muon hodoscope. The signals from
 muon-decay electrons and the evolution of the charge with water level are also studied. }

\begin{document}

\maketitle

\section{Introduction}
\vspace{-0.1cm}
The surface array of the Pierre Auger Observatory is a regular grid of
cylindrical water Cherenkov tanks with a 1500 m spacing that samples
the shower particles at the ground level
\cite{EANIM}.  The particles reaching the ground are mainly photons,
electrons and muons, with mean energies around 10 MeV
for photons and electrons and about 1 GeV for muons.
Cherenkov radiation is emitted in the water by the electrons and the muons as well as by electrons produced by photons converted via Compton scattering
and pair production.
In each detector the water is highly purified and contained in a Tyvek\textsuperscript{\textregistered} bag with a
diffusively reflective white surface to maximize the path length of Cherenkov photons and thus their chance to be collected. The light is
detected by three 9 inch photomultiplier tubes (PMTs) viewing the tank
from the top. The PMT signals are processed and digitised by 40 MHz Flash Analog-to-Digital Converters \cite{EANIM}.\\
The full simulation chain of the tank response, from the produced light to the digitized signals, is performed with a
dedicated program based on GEANT4  \cite{g4_2}.
Here we describe the implementation of the code and we will validate this tool by
comparing the simulated response of an Auger tank to experimental data on
muons crossing at different incident angles and water levels, and to electrons from muons stopping in the tank.
\section{Tank simulation framework}
The Auger tank simulation, which is a part of the  Auger DPA Offline package \cite{dpa}, is based on the well
established GEANT4 package. A dedicated module, called G4Fast, has been implemented to reduce the computing time.
 This module produces Cherenkov photons along
the path of the injected particle and tracks them through the water until they are absorbed or they reach the
active photocathode area of a PMT.
The output is the number of photoelectrons as a function of time which is
then processed by a different module simulating the PMTs and electronics response.\\
 Properties of a typical Auger water detector, such as the geometry of the
tank and of PMTs, materials properties, etc.\cite{EANIM} are included in the simulation, but
it is impractical to establish with precision
the details of the individual properties of all the PMTs for each tank.
Instead we use realistic average values given by their manufacturer:
Photocathode Area = 426 $cm^{2}$, Maximum Quantum Efficiency (QE) = 0.24, Collection Efficiency (CE) = 0.7. It must be noted that while
these parameters directly influence the number of photoelectrons produced by each PMT,
the values are not crucial for understanding the responses of the detectors. Instead, the detectors are continuously
calibrated with atmospheric muons: the measured signals from showers are given in units of the charge of a vertical muon crossing
the center of the tank \cite{calibnim} ($Q_{VEM}$ or VEM).
The simulated signals consequently also are given in VEM units.\\
We also use the maximum water absorption
length ($L$) = 100 m and the maximum Tyvek\textsuperscript{\textregistered} reflectivity ($R$) = 0.940; these choices are discussed below.
\vspace{-0.2cm}
\section{Response of the tank for different water levels}
\begin{figure}[t]
\vspace{-0.4cm}
    \centering
        \includegraphics[width=7.7cm,height=6cm]{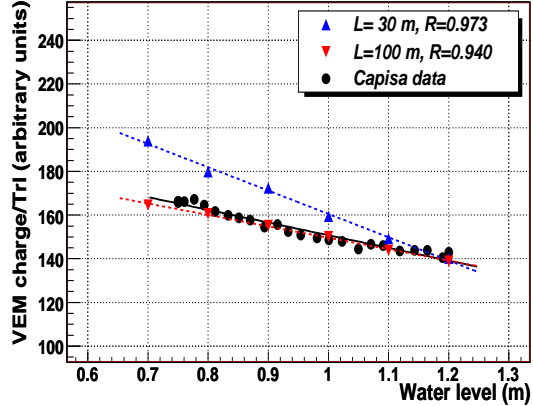}
\vspace{-0.8cm}
    \caption{VEM charge relative to the tracklength as a function of water level}
    \label{fig:waterlevel}
\vspace{-0.5cm}
\end{figure}

\vspace{-0.2cm}
The water and Tyvek\textsuperscript{\textregistered} parameters, $L$ and $R$, influence the propagation of the light in tank
and hence how the signal decreases with time almost exponentially after the first reflections. The chosen values
of the parameters are such that the measured decay time is reproduced well by simulations of simulated vertical muons. In fact, different pairs
of values for $L$ and $R$ could reproduce equally well the experimental decay time. We present in this section a study meant
not only to validate the simulation but also to disentangle $L$ and $R$.\\
An experiment was performed at the Auger site in a tank instrumented with scintillators to select vertical muons, where the
water level was decreased over a week from 120 cm to 75 cm. The charge deposited by the vertical muons was measured
for every 2 cm drop in water level. Simulations were performed using G4Fast with the water/Tyvek\textsuperscript{\textregistered}  parameters
$L$, $R$ given in the previous section and with another set, where the Tyvek\textsuperscript{\textregistered} quality was
improved and the water made more attenuating ($L$ = 30 m and $R$ = 0.973).\\
Working with different water levels changes the relative influence of water and Tyvek\textsuperscript{\textregistered}.
Less water means a reduced volume where photons impact more often on the Tyvek\textsuperscript{\textregistered}: the importance
of the reflectivity is expected to increase as the water level decreases.
The main effect of less water is to decrease the track length of vertical muons, and so the deposited charge. However, at the
same time, the charge per unit of length increases, and we expect an enhanced effect for the larger $R$.  The VEM charge,
normalized to the tracklength, versus water level is shown in Fig \ref{fig:waterlevel} for data and for simulation with the
two different sets of parameters. The expected effect of Tyvek\textsuperscript{\textregistered} is clearly demonstrated and
the data are completely consistent with the chosen parameters.
\vspace{-0.2cm}
\section{Response of the tank to vertical and omni-directional muons}
\begin{figure}[H]
\vspace{-0.7cm}
\begin{center}
\noindent
\hspace{-0.5cm}
\includegraphics [width=7.5cm,height=5.5cm]{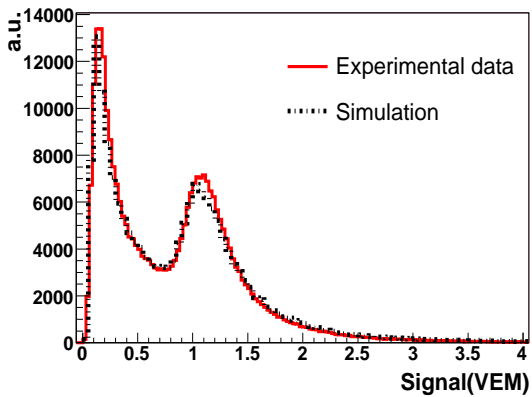}
\end{center}
\vspace{-0.8cm}
\caption{Charge distribution (VEM units) for omni-directional particles. Full line corresponds to data and dashed one to simulation.}
 \label{fig:charge}
\end{figure}

\vspace{-0.2cm}
The basic calibration information is the charge deposited by vertical and
central throughgoing muons. In this section, we compare the simulation
with vertical muon data as a first test of the simulation.\\
The water tank, in its normal configuration, has no way to select only vertical muons: however the distribution
of charges deposited by omni-directional muons has a peak which is well correlated with the VEM charge~\cite{calibnim}. The
peak is at 1.09 VEM, measured in an Auger tank instrumented with a muon hodoscope~\cite{orsay_exp}. This
ratio is an essential parameter to be reproduced by the simulations.
Using G4Fast, we simulate vertical muons (i.e. passing through the center of the tank and crossing the entire
volume of water) as well as omni-directional muons.  We use a realistic spectrum of
multi-directional electrons, muons and photons with the energy spectrum from reference \cite{muon_energy_spectrum}. We
assume, as an approximation, the same zenith angle distribution,
$f(\theta) = \cos^2(\theta)\sin(\theta)$, for all particle types.
We use a low-threshold trigger requiring a 3-fold coincidence over
0.15 $I^{peak}_{VEM}$ in each PMT as in real data (being $I^{peak}_{VEM}$ the
average of the peak in the pulse produced by vertical muons).
 Fig. \ref{fig:charge} shows the comparison of the experimental and simulated charge distribution for the omni-directional
 muons in units of VEM: the simulation reproduces the data well. The position of the peak is found
 to be 1.09 VEM, as in data.
\vspace{-0.6cm}
\section{Response of the tank to inclined muons}
\begin{figure}[t]
    \centering
        \includegraphics[width=5.1cm]{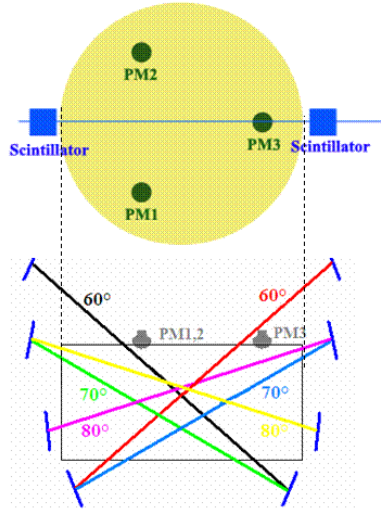}
\vspace{-0.5cm}
    \caption{Set-up of the Orsay tank.}
    \label{fig:config}
\vspace{-0.3cm}
\end{figure}

\begin{figure}[H]
\vspace{-0.3cm}
    \centering
        \includegraphics[width=7.4cm]{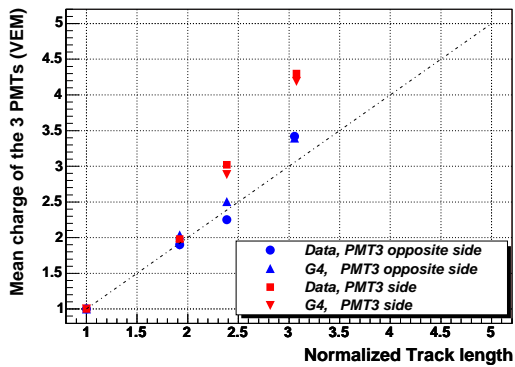}
\vspace{-0.9cm}
    \caption{Mean charge of the 3 PMTs as a function of tracklength in tank (both quantities are normalized to the VEM)}
    \label{fig:inclined}
\end{figure}

\vspace{-0.1cm}
To validate the simulations of inclined muons
we compared the simulated values of the recorded charge with
measurements from a test tank in Orsay, similar to the Auger ones,
where signals from atmospheric muons could be recorded at
different zenith angles. Two movable scintillators were placed at the side of the
tank, triggering on muons arriving with different angles, disentangled from correlated shower events by means of
proper timing ~\cite{orsay_exp}. The scintillators were located for each incident angle in two opposite positions
as shown in Fig.~\ref{fig:config}. As the average energy of the muons increases
with zenith angle, we use the energy
parameterisation given in \cite{muon_energy_spectrum}. The simulated and experimental charges are plotted versus the muon tracklength
in Fig.~\ref{fig:inclined}.
A deviation from linear behaviour is observed in the data as the zenith angle
increases, due to the appearance of direct non-reflected light, in particular when
muons cross close to the PMT. This behavior is reproduced well (within 10\%) by the simulation.
\vspace{-0.5cm}
\section{Response of the tank to electrons from muon decay}
\vspace{-0.2cm}
Muons decaying in the tank produce electrons with a well known energy spectrum, the Michel spectrum, with an end point at 53 MeV and
an average value of 37 MeV. The measurement of the Cherenkov light produced by Michel electrons provides a reference point for
the tank response to low energy electrons.
A dedicated trigger was implemented to select muon decay events and allowed measurement of the ratio between the Michel spectrum peak and
that of vertical throughgoing muons: the average value for the 230 tanks in operation at that time was found to be 0.13 with a tank-to-tank
spread of 0.01.  A study with G4Fast used the same algorithm as used in the data to select the decaying muons. Crossing muons were
generated with an appropriate angular distribution and energy spectrum~\cite{muon_energy_spectrum}.
The simulated ratio between the electron and muon charge, $0.13\pm0.01$, agrees well with the measured one.
\begin{figure}[H]
\vspace{-0.5cm}
    \centering
        \includegraphics[width=7.8cm,height=6cm]{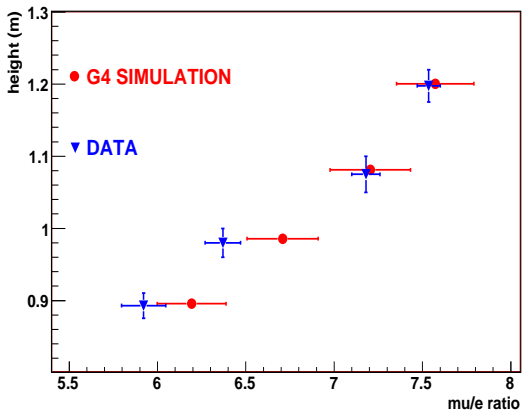}
\vspace{-0.8cm}
    \caption{Evolution of muon to electron charge with the water level}
    \label{fig:muen}
\end{figure}
\vspace{-0.2cm}
Michel electrons are absorbed in less than 25 cm and therefore are almost insensitive to a change in the water level,
while vertical muons produce a signal mostly proportional to it.
A linear dependence of the
ratio between electrons and muons versus water level is expected. An experiment was performed to test the influence of water
loss: a tank was slowly drained starting from its normal level at $1.2$ m and finishing at $0.895$ m ~\cite{muondecay}. Four
ratios were obtained during
this process, shown in Fig.~\ref{fig:muen} as triangles. Simulations were carried out for the same
water levels and the $e/\mu$ ratios computed as described previously. The compatibility
between the simulation points and the data is visible in the same figure.

\section{Conclusions}
\vspace{-0.2cm}
The simulation of the Auger water Cherenkov tank is accomplished by a module based on GEANT4,
designed to reduce the computing time (so called G4Fast). We have described a variety of tests of the simulation versus data:

- Vertical muons. The ratio between the VEM charge (the basis of the calibration of the Auger
surface detector) and the average charge detected for omnidirectional muons has been measured in a dedicated experiment to be
0.92. The same value is found using G4Fast simulations of both vertical and multidirectional muons, and their spectrum also is reproduced well.

- Inclined muons. Due to their increased pathlengths in the tank, inclined muons yield a larger charge. The behaviour of the signal
versus tracklength has been measured in an ad hoc experiment: the simulated charge response for different muon directions is
well represented by G4Fast, including the effects of direct light on the PMTs.

- Electrons from muon decay. We have measured and simulated the ratio between the charge peak from Michel electrons
and the VEM peak. The average experimental ratio (0.13, with a tank-to-tank dispersion of 0.01) is reproduced well by G4Fast.
Good agreement is found as well for the evolution of this ratio with changes in water level.
\bibliography{icrc0300}

\begin{thebibliography}{1}

\bibitem{EANIM}
J.~{Abraham et al. [Pierre Auger Collaboration]}.
\newblock {\em Nucl. Instr. Meth. Phys. Res. A}, 523:50, 2004.

\bibitem{g4_2}
J.~{Allison et al.}
\newblock {\em IEEE Transactions on Nuclear Science}, 53:270, 2006.

\bibitem{dpa}
S.~{Argiro et al. [Pierre Auger Collaboration]}.
\newblock {\em Proc. 29th ICRC}, 8:343, 2005.

\bibitem{calibnim}
X.~{Bertou et al.}
\newblock {\em Nucl. Instrum. Meth. Phys. Res. A}, 568:839, 2006.

\bibitem{orsay_exp}
M.~{Aglietta et al. [Pierre Auger Collaboration]}.
\newblock {\em Proc. 29th ICRC}, 7:83, 2005.

\bibitem{muon_energy_spectrum}
D.~{Reyna}.
\newblock {\em hep-ph/0604145}, 2006.

\bibitem{muondecay}
P.~{Allison et al. [Pierre Auger Collaboration]}.
\newblock {\em Proc. 29th ICRC}, 8:299, 2005.

\end{thebibliography}
\bibliographystyle{unsrt}
\end{document}